\documentclass[preprint,showpacs,preprintnumbers,amsmath,amssymb]{revtex4}

\def\ube13{UBe$\rm_{13}$}

\def\bi2212{Bi$\rm_2$Sr$\rm_2$CaCu$\rm_2$O$\rm_8$}
\def\ybi2212{Bi$\rm_2$Sr$\rm_2$YCu$\rm_2$O$\rm_8$}
\def\ycabi2212{Bi$\rm_2$Sr$\rm_2$Ca$\rm_{1-x}$Y$\rm_x$Cu$\rm_2$O$\rm_{8+\delta}$}

\def\y65cabi2212{Bi$\rm_2$Sr$\rm_2$Ca$\rm_{0.35}$Y$\rm_{0.65}$Cu$\rm_2$O$\rm_{8+\delta}$}

\def\Co{CeCoIn$_5$}

\usepackage{graphicx}

\begin{document}

\title{Anisotropy of Thermal Conductivity and Possible Signature of the Fulde-Ferrell-Larkin-Ovchinnikov state in CeCoIn$_5$}
\author{C. Capan,$^1$ A. Bianchi,$^{1,\ast}$ R. Movshovich,$^1$ A.~D.~Christianson,$^2$ A. Malinowski,$^1$ M. F. Hundley,$^1$ A. Lacerda,$^2$ P. G. Pagliuso,$^3$ J. L. Sarrao$^1$}
\affiliation{$^1$Los Alamos National Laboratory, Los Alamos, New Mexico 87545\\
             $^2$National High Magnetic Field Laboratory, Los Alamos, New Mexico 87545\\
         $^3$Instituto de Fisica Gleb Wataghin, UNICAMP, 13083-970, Campinas, Brazil}
\date{\today}
\begin{abstract}
We have measured the thermal conductivity of the heavy-fermion superconductor CeCoIn$_5$ in the
vicinity of the upper critical field, with the magnetic field perpendicular to the c axis. Thermal
conductivity displays a discontinuous jump at the superconducting phase boundary below critical
temperature $T_0 \approx 1$ K, indicating a change from a second to first order transition and
confirming the recent results of specific heat measurements on \Co. In addition, the thermal
conductivity data as a function of field display a kink at a field $H_k$ below the superconducting
critical field, which closely coincides with the recently discovered anomaly in specific heat,
tentatively identified with the appearance of the spatially inhomogeneous
Fulde-Ferrell-Larkin-Ovchinnikov (FFLO) superconducting state. Our results indicate that the
thermal conductivity is enhanced within the FFLO state, and call for further theoretical
investigations of the order parameter's real space structure (and, in particular, the structure of
vortices) and of the thermal transport within the inhomogeneous FFLO state.
\end{abstract}


\maketitle

\section{Introduction}
Over the last several years there has been renewed interest in the spatially inhomogeneuos
Fulde-Ferrell-Larkin-Ovchinnikov(FFLO) state. The FFLO state was predicted as early as the
mid-1960s~\cite{fulde-ferrell:pr-64,larkin-ovchinnikov:jetp-64} to occur in a clean Type II
superconductor in high magnetic fields, when the Zeeman energy becomes comparable to the
condensation energy. Then, Pauli limiting~\cite{clogston:prl-62} plays an important role in
defining both the superconducting critical field $H_{c2}$ and the temperature $T_0$ below which the
FFLO state is expected to appear~\cite{gruenberg:prl-66}. Within the FFLO state spin up and spin
down electrons of a spin-singlet superconductor can only stay bound if the Cooper pair has a finite
momentum. As a result, the FFLO state is formed with a spatially oscillating order parameter. The
exact description of the corresponding phase diagram for both s- and d-wave
superconductors,~\cite{houzet:prb-01,agterberg:jpcm-01,tachiki:zfpb-96,combescot:cond-mat-03} as
well as the stable spatial structures in 2D and 3D in the presence of
vortices~\cite{buzdin:plA-96,houzet:prb-01,shimahara:jpsj-97,klein:jltp-00}, are subjects of
intense theoretical investigations.

In spite of the straightforward nature of the theoretical prediction, the experimental observation
of the FFLO state has turned out to be a difficult task. In fact, very few superconductors fulfil
the necessary conditions for the formation of an FFLO state. The relative importance of the Pauli
and orbital limiting can be described by the so-called Maki parameter $\alpha = \sqrt 2
{H_{c2}^0\over H_p}$. $H_{c2}^0$ is the orbital limiting field due to the kinetic energy of the
superconducting currents around the vortex cores, commonly derived from the slope of the
experimentally determined $H-T$ phase boundary at $T_c$ as $H_{c2}^0 = 0.7 {d H_{c2} \over d
T}|_{T_c}$~\cite{werthammer:pr-66}. $H_p = \sqrt 2 \Delta_0 / g  \mu _B$ is the Pauli limiting
field due to the potential energy of electrons' spins (Zeeman energy). Here $\Delta_0$ is the zero
temperature value of the superconducting gap, $g$ is the electron's effective g-factor, and $\mu_B$
is the electron's Bohr magneton~\cite{clogston:prl-62}. Within the calculation of
Ref.~\cite{gruenberg:prl-66}, $\alpha$ must be greater than 1.8 for the FFLO state to be realized.

There are several classes of materials that are traditionally thought of as potential candidates
for the formation of the FFLO states. These include low dimensional organic superconductors and
heavy fermion superconductors. The low dimensional organic superconductors are promising because
when the field is applied within the conducting planes of a 2D superconductor, the orbital limiting
is suppressed entirely, as the diamagnetic screening currents can only flow within the plane. In
such case Pauli limiting determines the critical field $H_{c2} = H_p$, the Maki parameter $\alpha =
\infty$, and the FFLO state should be stabilized below the critical temperature $T_0 \approx 0.55
T_c$ for magnetic field close to $H_{c2}$. This straightforward prediction led to a number of
experimental investigations of the superconducting properties of lower-dimensional organic
superconductors. Several investigators suggested the existence of FFLO states, e.g. based on the
superconducting phase diagram~\cite{singleton:jpcm-00} or the magneto-thermal transport
properties~\cite{tanatar:prb-02}.

Heavy fermion superconductors are also attractive because of their potentially large values of the
Maki parameter. Here the heavy electron masses lead to low Fermi velocities of the quasiparticles
and in turn to relatively ineffective orbital limiting, or large $H_{c2}^0$. For this reason heavy
fermion materials have had their deserved share of attention, and the features in the magnetization
of CeRu$_2$~\cite{huxley:jpcm-93} and the phase diagram of UPd$_2$Al$_3$~\cite{gloos:prl-93} were
taken as possible signatures of the FFLO states.

However, and not for the lack of effort, there is to date no accepted definitive proof of the
existence of the FFLO state in any of the systems described above. In this paper we describe the
magneto-thermal studies of a new and very strong candidate to possess the FFLO state, the
heavy-fermion superconductor \Co.

CeCoIn$_5$ is a clean d-wave superconductor,~\cite{petrovic:jpcm-01,movshovich:prl-01,izawa:prl-01}
with a T$_c$ of 2.3K, the highest among the Ce-based heavy fermions. It exhibits a layered
structure of alternating CeIn$_3$ and CoIn$_2$ planes, suggesting a possible quasi-2D electronic
nature of this compound. This is supported by the experimental observation of a quasi-cylindrical
sheet in the Fermi surface of \Co\ via de Haas-van Alphen studies~\cite{hall:prb-01}. The estimated
Maki parameter $\alpha$ is about 3.6 for the field perpendicular to the CeIn$_3$ planes ($H
\parallel c$)~\cite{bianchi:prl-02}, and close to 4.5 for the field-in-the-plane orientation ($H
\perp c$)~\cite{won:cond-mat-03}. Thus, CeCoIn$_5$ is a good candidate for the formation of an FFLO
state. As new results accumulate, there is a growing evidence that the FFLO state may indeed be
realized in CeCoIn$_5$. First, the superconducting phase transition at low temperatures becomes
first order, which is manifested by the sharp specific heat anomaly at $T_c$ in the specific heat
for both $H \parallel c$~\cite{bianchi:prl-02} and $H \perp c$~\cite{bianchi:prl-03a} orientations.
In addition, steps in magnetostriction~\cite{bianchi:prl-02,takeuchi:jpcm-2002} and in
magnetization~\cite{tayama:prb-02,murphy:prb-02}, and a step in the thermal conductivity for $H
\parallel c$~\cite{izawa:prl-01} are observed. The change of the superconducting anomaly from second to first
order~\cite{bianchi:prl-02} was interpreted as a realization of the Maki scenario, which attributes
this change to a strong Pauli limiting effect in a Type II
superconductor~\cite{maki:ptp-64,maki:pr-66}. When the field was applied within the $a-b$ plane ($H
\perp c$), a second anomaly in the specific heat was observed within the superconducting
state~\cite{bianchi:prl-03a,radovan:nature-03}, indicating a phase transition into a new
superconducting state, tentatively identified as the FFLO state in \Co. In addition, steps in
magnetization of \Co\ were observed by Radovan {\it et al.}~\cite{radovan:nature-03}, and were
interpreted as an indication of the multi-quantum vortices expected under certain circumstances for
the 2D superconductors within the FFLO state~\cite{buzdin:plA-96,shimahara:jpsj-97,klein:jltp-00}.
The validity of such interpretation is at present under
debate~\cite{movshovich:nature-04,radovan:nature-04}. On the theoretical front, recent analysis of
a linearly increasing H$_{c2}$ at the lowest temperatures~\cite{won:cond-mat-03} suggested that
this too can be accounted for within an FFLO scenario for \Co.

While these results make the FFLO scenario a very appealing one for CeCoIn$_5$, there is no clear
evidence so far for spatially inhomogeneous superconductivity in the second low temperature phase.
A recent study revealed an increased penetration depth at the lower transition, which was
interpreted as a decrease of the superfluid density due to the formation of the FFLO
state~\cite{martin:cond-mat-03}. An ultrasound investigation of the high field state revealed the
decrease of the sound velocity from that in the vortex state, which was also presented in support
of the FFLO nature of that state~\cite{watanabe:condmat-03}.

\begin{figure}
\resizebox{!}{0.9\textwidth}{\includegraphics{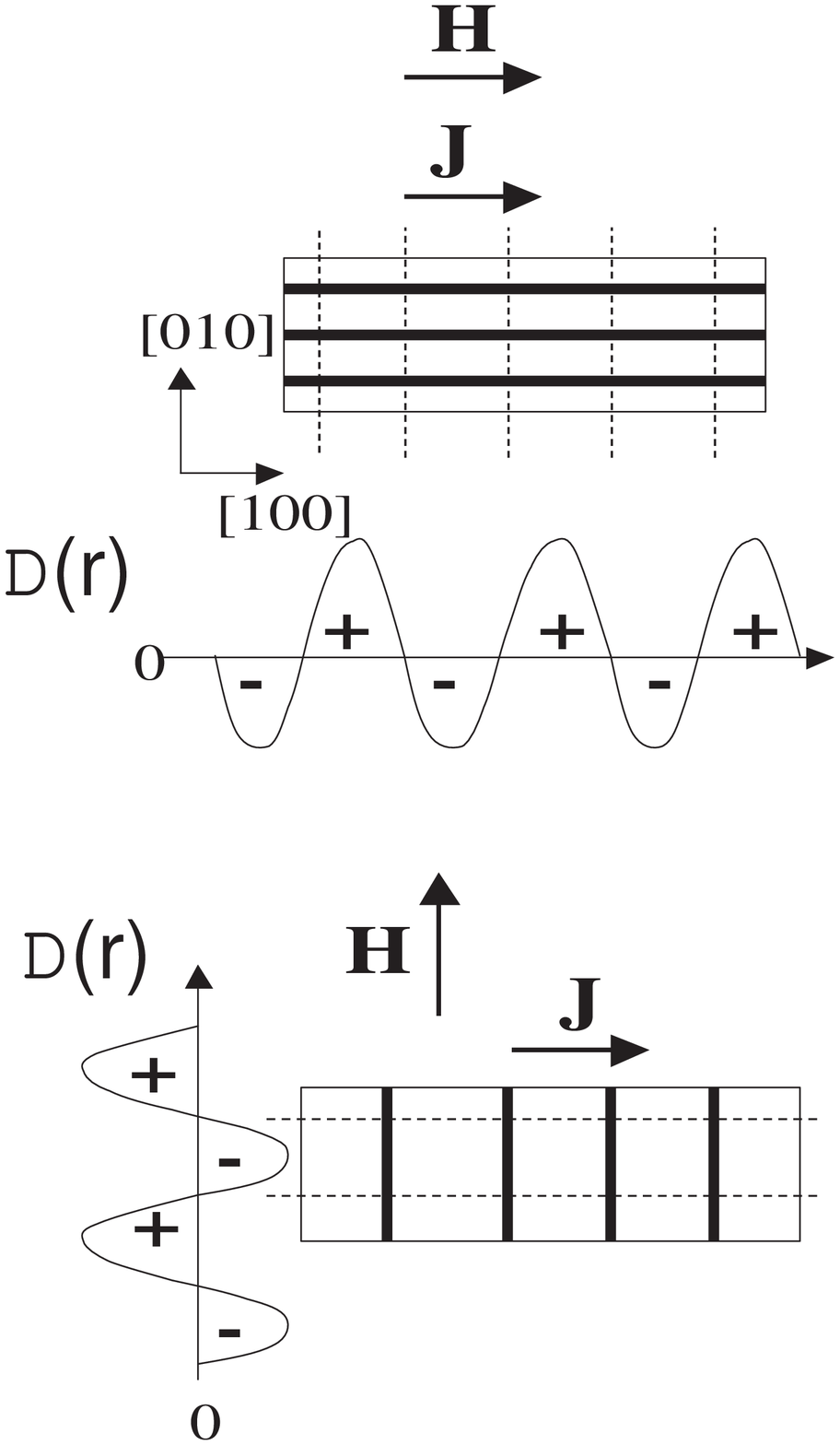}} \caption{ Illustration of the vortex
structure (solid lines) and the FFLO modulation (dashed lines) with the field parallel  (top) and
perpendicular (bottom) to the heat current.} \label{FFLO-illustration}
\end{figure}

Here we present our results of magneto-thermal transport measurements in \Co\ with the field
applied within the CeIn$_3$ planes ($H \perp c$). The LO structure, which emerged from the original
theoretical work~\cite{larkin-ovchinnikov:jetp-64}, is a collections of periodically spaced planes
of nodes of the superconducting order parameter that are perpendicular to the direction of the
applied field. The LO order parameter is described as $\psi(\vec r) = \psi_0 \cos (\vec q \vec r)$,
oscillating in space along the direction of vector $\vec q \parallel H$ as illustrated in
Fig.~\ref{FFLO-illustration}. In recent years, thermal conductivity was used effectively to probe
the anisotropy of the order parameter in unconventional superconductors, specifically, the
structure of the nodes in $k$-space. This is due to the fact that normal quasi-particles, which are
easily excited along the nodal directions, do carry heat, whereas the superconducting background
does not~\cite{aubin:prl-97,izawa:prl-01,vekhter:phC-00}. The inherent anisotropy of the LO state
makes thermal conductivity an attractive tool for identifying it. For the vast majority of clean
superconductors thermal conductivity drops as the sample enters the superconducting state due to
the opening of the superconducting gap over the entire Fermi surface, and the resulting rapid
decrease of the number of normal quasiparticles which carry heat. One would then expect the thermal
conductivity within the nodal planes to be higher than in the rest of the sample. Thermal
conductivity would then be larger when the heat current $\vec Q$ flows along the nodal planes, and
perpendicular to the applied magnetic field ($\vec Q \perp H$), than when the heat flow is parallel
to the magnetic field ($\vec Q \parallel H$). Note that this anisotropy is of the opposite sign
from that due to the vortices, with higher thermal conductivity along the vortices ($\vec Q
\parallel H$),~\cite{maki:pr-67} and therefore the two contributions should be easily
differentiated, especially if the contribution due to the 2D planes turns out to dominate that from
the 1D vortices. This picture unfortunately turns out to be too simplistic for the case of \Co\ and
is complicated by several effects described below.

Motivated by the idea described above, we measured thermal conductivity in CeCoIn$_5$ at low
temperatures, in the vicinity of the upper critical field, with the magnetic field oriented
in-plane, using a dilution refrigerator in the 20 T magnet of the NHMFL facility at LANL. The
sample, a needle-like single crystal with  dimensions of(2.18 mm, 0.28 mm, 0.064 mm), was
flux-grown at LANL as described in Ref.~\cite{petrovic:jpcm-01}. After a chemical etch and
polishing to remove the residual free indium, the sample had resistivity of 3.2 $\rm \mu \Omega -
cm$ at 4.2 K and a RRR of ${{\rho (300 K)}\over {\rho(4.2 K)}} = 9.4$. The experimental set-up for
thermal conductivity consists of a heater attached to one end of the sample, and two RuO$_2$
thermometers in thermal contact with the sample at two points along its length. A DC heat current
flows along the [100] direction of the sample (along its longest dimension). The resulting
temperatures of both thermometers are measured using an LR700 resistance bridge. The measurements
were performed with magnetic field applied parallel  and perpendicular ($H \parallel$ [010]) to the
heat current. The two thermometers were calibrated at each field against a reference thermometer
placed in a field-free region.

\section{Thermal conductivity in the normal state and zero field}

\begin{figure}
\resizebox{!}{0.7\textwidth}{\includegraphics{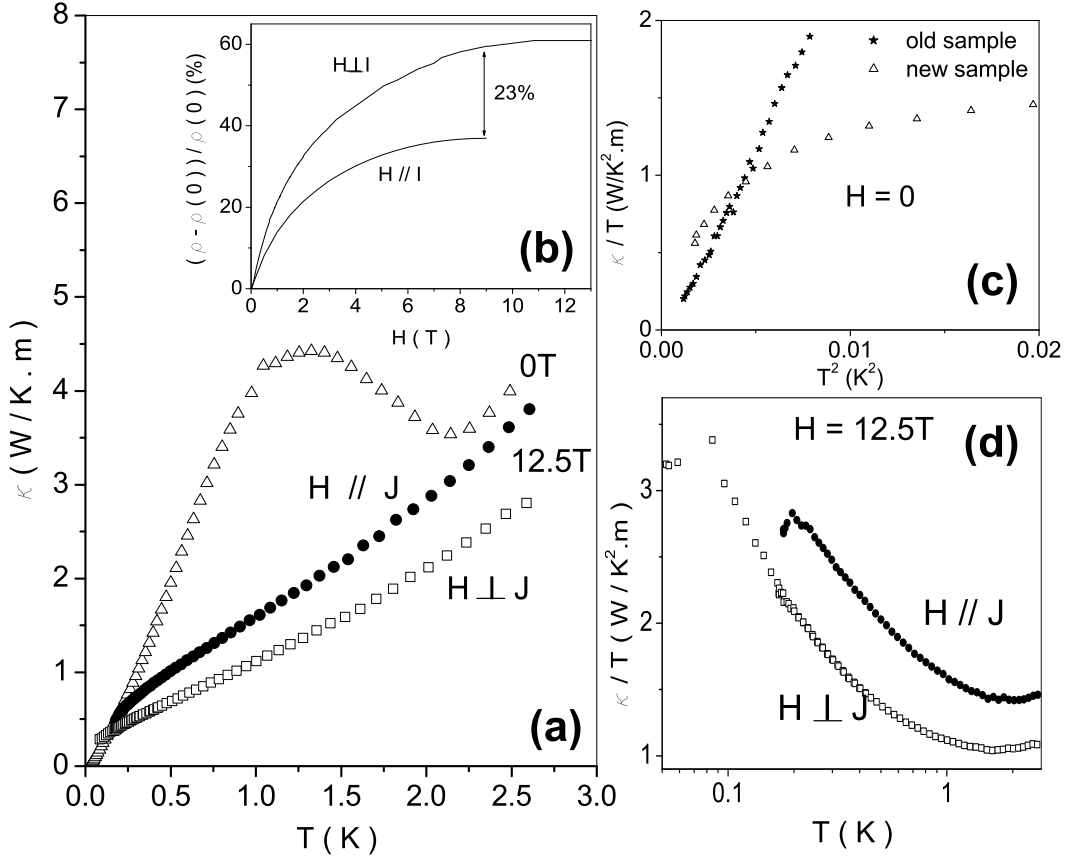}} \caption{(a): Thermal conductivity
$\kappa$ vs. temperature of CeCoIn$_5$ at zero field and 12.5 T, with the field parallel
($\bullet$) and  perpendicular ($\Box$) to the heat current. Inset (b): In-plane magnetoresistance
of CeCoIn$_5$ with field perpendicular (top curve) and parallel (bottom curve) to the current. (c):
Thermal conductivity of the present sample ($\triangle$) and the sample used in Ref. [18]
($\star$), normalized to the value at $T_c$ of $\kappa = 2$ $\rm W/K m$. (d) Thermal conductivity
divided by temperature for the field of 12 T parallel ($\bullet$) and perpendicular ($\Box$) to the
heat current, showing diverging $\kappa/T$ as $T \rightarrow 0$.} \label{k-anisotr}
\end{figure}

Fig.~\ref{k-anisotr} shows the thermal conductivity of \Co\ as a function of temperature up to 2.5
K at zero field and at 12.5 T, for field oriented parallel and perpendicular to the heat current.
One notices a substantial drop in thermal conductivity induced by the applied field above T$_c$ as
well as a significant difference between the two field orientations. It is possible to account for
the normal state anisotropy by considering magnetoresistance of \Co, displayed in the inset of
Fig.~\ref{k-anisotr}. The longitudinal and transverse magnetoresistances at 2.5K and 9T, with an
in-plane current, are 37\% and 60\% respectively. This gives a difference in magnetoresistance
$(\rho_{\perp}-\rho_{\parallel})/\rho(0)$ of 23\% between the field oriented parallel and
perpendicular to the electrical current. If we determine the difference of thermal conductivity
between both field orientations, $(\kappa_{\parallel}-\kappa_{\perp})/\kappa(0)$ at 2.5 K and 12.5
T we find a value of 21\%, very close to the anisotropy of magnetoresistance. Thus, the anisotropy
of the heat transport in the normal state can be accounted for by the difference in the
quasiparticle scattering for the two field orientations. This is not surprising because the
quasiparticle contribution dominates the heat transport in \Co~\cite{movshovich:prl-01}. Therefore,
in order to highlight the differences between the superconducting states for the two orientations
of the magnetic field studied, in what follows, we often present the thermal conductivity data
scaled by the values in the normal state at 12.5 T.

The normal state thermal conductivity for the field of 12.5 T, just above the superconducting
critical field of 12 T, is displayed in Fig.~\ref{k-anisotr}(d) as $\kappa \over T$ vs. $T$ on a
log-log plot. $\kappa \over T$ appears to diverge, reflecting the possible presence of the quantum
critical point (QCP) in \Co, suggested by both specific heat and resistivity
measurements~\cite{capan:tbp-04}. Similar behavior with the QCP lying very close to the
superconducting critical field was observed for $H \parallel$
[001]~\cite{sidorov:prl-02,paglione:prl-03,bianchi:prl-03b}.

It is interesting to compare the zero-field thermal conductivity measurements on the present sample
to the previously published data~\cite{movshovich:prl-01}, in particular the zero-temperature limit
of $\kappa \over T$. The zero-field data are displayed in Fig.~\ref{k-anisotr}(c) as $\kappa \over
T$ vs. $T^2$, and appear to extrapolate to the value of a few tenths of $\rm W / K m$, close to the
value obtained for the original sample. This agreement supports the existence of the universal,
independent of the impurity concentration, limit of $\kappa \over T$ in \Co, and the original
interpretation of the low temperature thermal transport reflecting the presence of lines of nodes
in the superconducting energy gap of \Co~\cite{movshovich:prl-01}.

\section{Thermal conductivity in the vortex state}

\begin{figure}
\resizebox{!}{0.7\textwidth}{\includegraphics{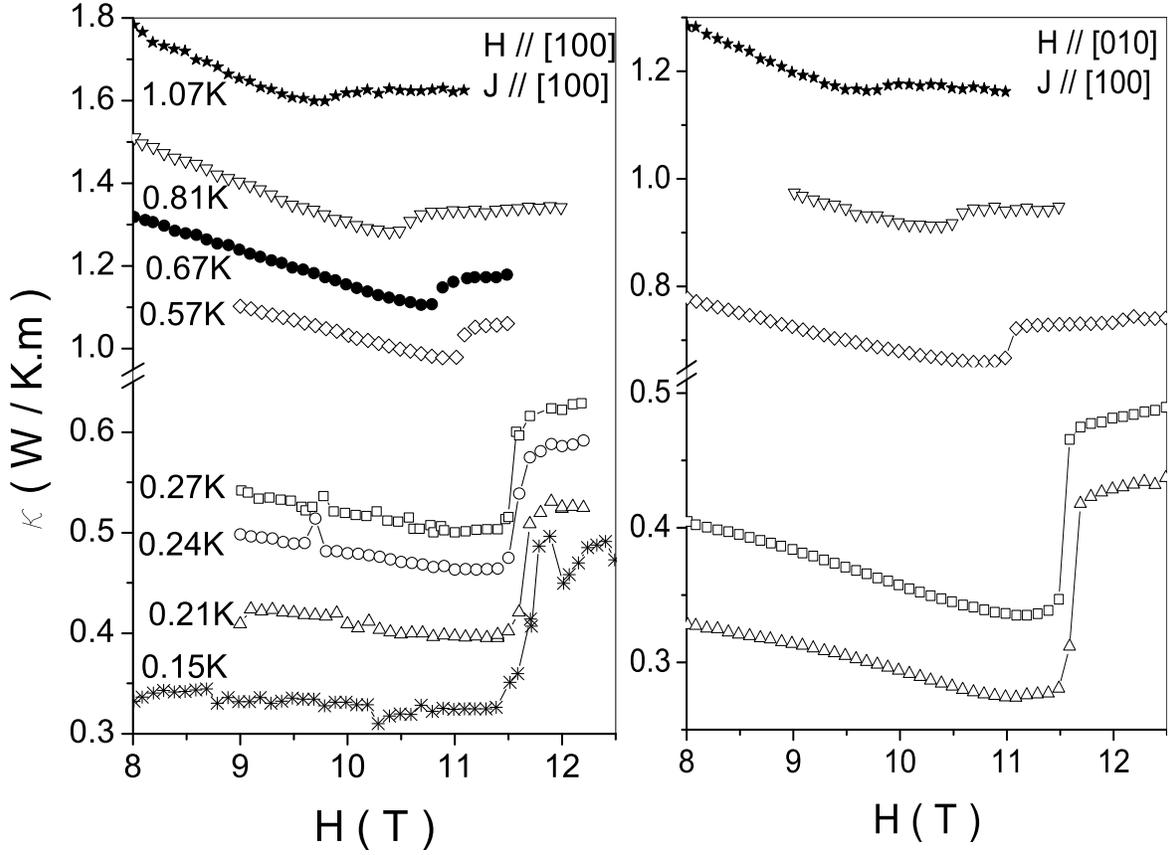}} \caption{ Thermal conductivity vs.
magnetic field of CeCoIn$_5$ between 8T and 12.5T. Left Panel: $H \parallel J$. Right Panel: $H
\perp J$. ($\star$) 1.07 K, ($\bigtriangledown$) 0.81 K, ($\bullet$) 0.67 K, ($\diamond$) 0.57 K,
($\Box$) 0.27 K, ($\circ$) 0.24 K, ($\bigtriangleup$) 0.21 K, ($\ast$) 0.15 K. } \label{k-vs-H}
\end{figure}

The thermal conductivity data in the low temperature - high field part of the phase diagram are
displayed in Fig.~\ref{k-vs-H}. The transition to the normal state is marked by a pronounced jump
in the thermal conductivity at the lowest temperatures. The jump in thermal conductivity confirms
the first order nature of the superconducting transition, reported previously on the basis of the
specific heat measurements~\cite{bianchi:prl-03a}. The first order nature of the superconducting
transition for $H \parallel$ [001] was deduced on the basis of the thermal conductivity
measurements by Izawa {\it et al.}~\cite{izawa:prl-01}.

\begin{figure}
\resizebox{!}{0.7\textwidth}{\includegraphics{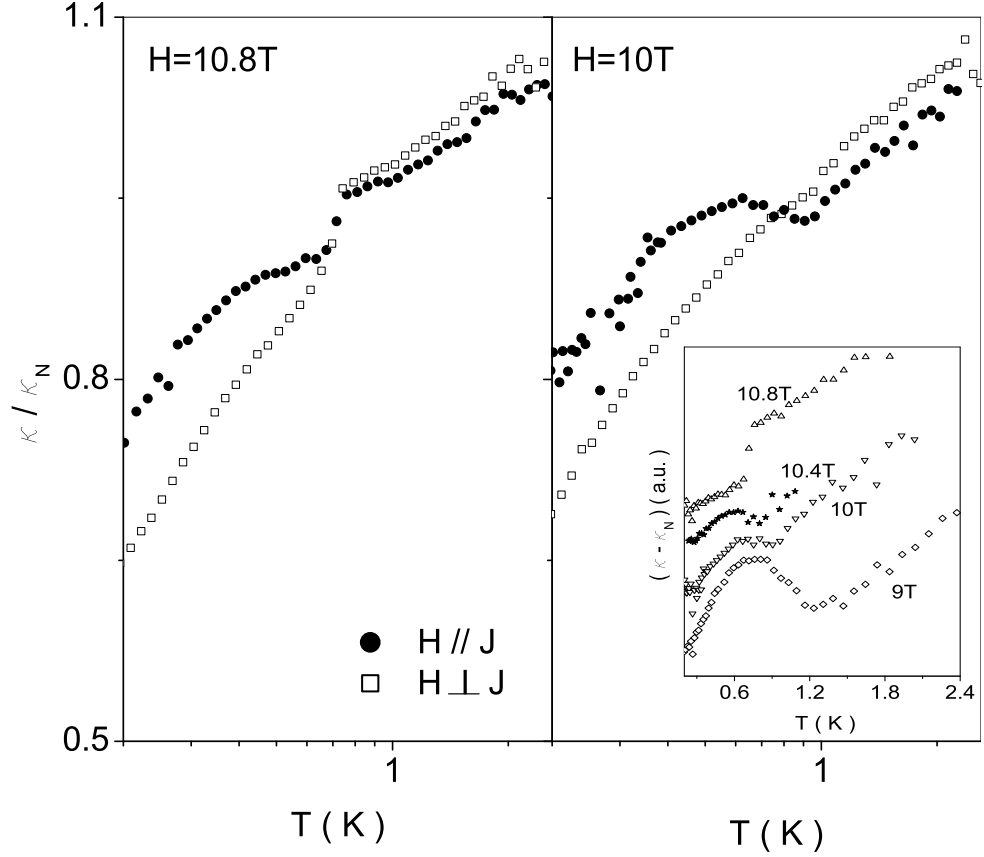}} \caption{Temperature dependence of
thermal conductivity $\kappa$ in CeCoIn$_5$ between 50 mK and 2.5 K. ($\bullet$) $H
\parallel J$; ($\Box$) $H \perp J$. Left panel: 10 T; right panel: 10.8 T; inset:
$\kappa-\kappa_N$ vs. temperature at 9 T ($\diamond$), 10 T ($\bigtriangledown$), 10.4 T ($\star$),
and 10.8 T ($\triangle$). The values of $\kappa$ at 12.5 T have been used as $\kappa_N$, and the
data for different fields have been shifted vertically for clarity.} \label{koverT-vs-T}
\end{figure}

The normalized thermal conductivity for both parallel and perpendicular field orientations is
depicted in Fig.~\ref{koverT-vs-T}(a,b) as a function of temperature for several values of field
between 50 mK and 2.5 K. The inset in Fig.~\ref{koverT-vs-T} shows the data after subtraction of
the normal state 12.5 T data. The enhancement of thermal conductivity below T$_c$ at zero field due
to an increased quasiparticle mean free path, shown in Fig.~\ref{k-anisotr}(a), can still be
clearly resolved at 9 T. An apparent small rise in thermal conductivity at the superconducting
transition at 10 T and 10.4 T is due to the competition between the increase of the thermal
conductivity in the normal state with increasing magnetic field and a drop in thermal conductivity
as the system becomes superconducting when magnetic field is swept (see Fig.~\ref{k-vs-H}). When
temperature is swept, the effect of the increasing $ \kappa$ in the normal state prevails at 10 T
and 10.4 T. The data for 10.8 T show a pronounced step (similar to the data for the field sweeps)
associated with the first order nature of the superconducting transition, which dominates at this
field.

\begin{figure}
\resizebox{!}{0.8\textwidth}{\includegraphics{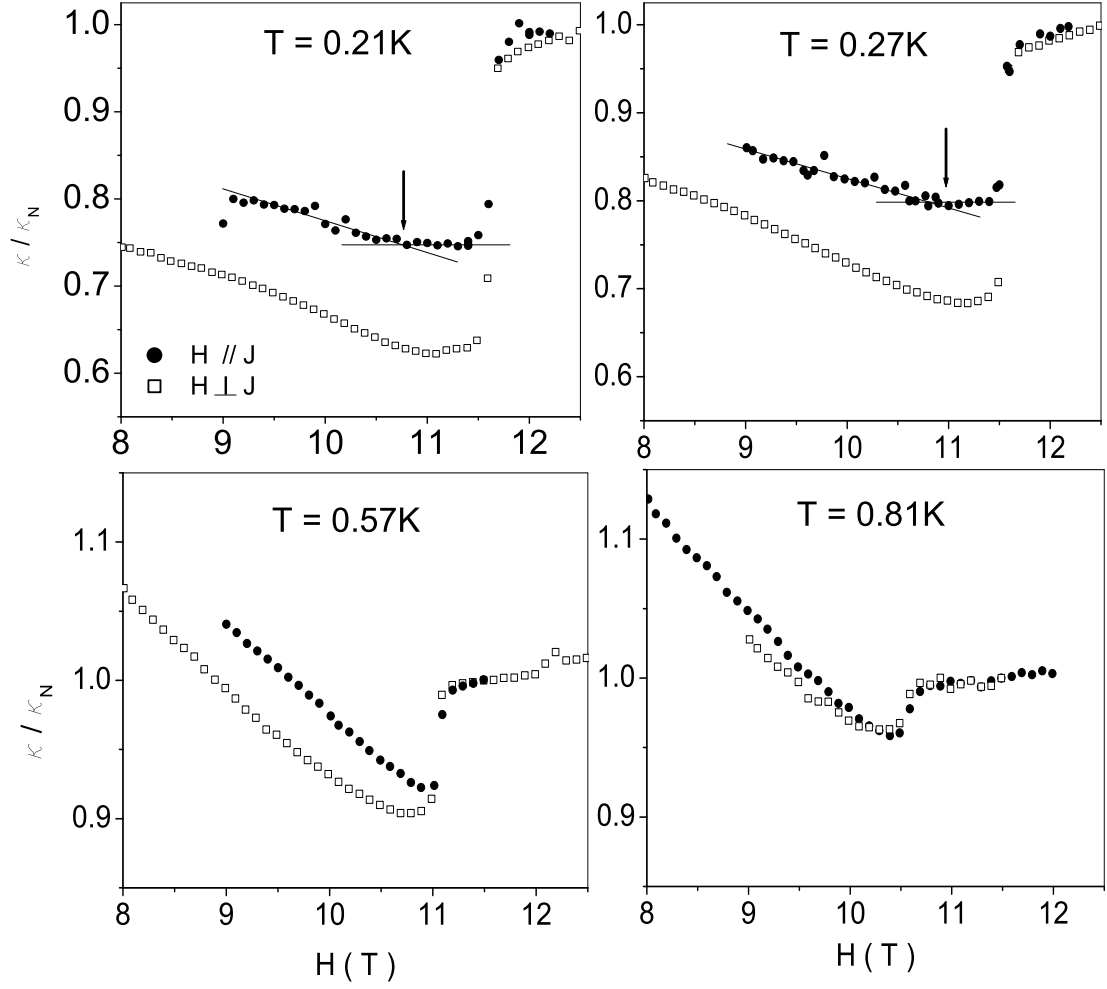}} \caption{ Normalized thermal
conductivity vs. magnetic field of CeCoIn$_5$ for $H \parallel J$ ($\bullet$) and $H \perp J$
($\Box$). Upper left: 0.21 K, upper right: 0.27 K, lower left: 0.57 K, lower right: 0.81 K. The
data for both orientations have been normalized by the corresponding normal state values at 12.5 T.
Arrows indicate positions of the kinks for $H \parallel J$.} \label{knorm-vs-H}
\end{figure}

Fig.~\ref{knorm-vs-H} displays the thermal conductivity normalized by the value in the normal state
at $T_c$ at several temperatures for the two field orientations between 8 T and 12 T. The thermal
conductivity decreases with increasing field in the mixed state and increases slightly in the
normal state. The absolute slope of the thermal conductivity vs. magnetic field in the vortex state
is reduced as the temperature is lowered. This is likely a result of the competition between the
effects of temperature and magnetic field. Magnetic field has two competing effects on the
quasiparticle heat transport. First is the Volovik effect, or doppler shift of the quasiparticle
energies, which results in the finite density of states in d-wave
superconductors~\cite{volovik:jetp-93,vekhter:prl-99}. This leads to the increase in thermal
conductivity with the increasing magnetic field. The second effect is due to the quasiparticle
scattering from vortices. As the field is increased, so is the number of scattering vortices, and
this has the effect of decreasing thermal conductivity in higher magnetic field. At high
temperature the number of quasiparticles is largely determined by temperature, and the contribution
from Volovik effect loses its significance. The vortex scattering effect then dominates thermal
transport, resulting in the decrease of thermal conductivity with magnetic field. On the other
hand, at low temperature the Volovik effect dominates the thermal broadening, and efficiently
competes with the reduction of thermal conductivity due to the vortex scattering. In \Co\ this
results in a slower decrease of thermal conductivity with increasing magnetic field at lower
temperatures, as displayed in Fig.~\ref{knorm-vs-H}. We do not reach the $\sqrt H$ regime observed
in YBCO ~\cite{chiao:prl-99} or BSCCO~\cite{aubin:prl-99} high temperature superconductors. Perhaps
\Co\ must be cooled to lower temperature to observe this field dependence due to its very small
impurity band width~\cite{movshovich:prl-01}. In the temperature and field range studied, the
quasiparticle scattering in \Co\ appears to be dominated by vortices, rather than the impurities.

The overall values of thermal conductivity for $H \parallel J$ orientation are higher than that for
$H\perp J$. At 0.21 K and 0.27 K the difference is close to 12\%, and at higher temperatures it is
significantly reduced (only 4\% at 0.57 K and 2\% at 0.81 K). This anisotropy of thermal
conductivity is due to the vortex scattering of the quasiparticles. More precisely, in a
semi-classical approach the scattering off the vortex is maximal when the quasiparticle velocity is
perpendicular to it, resulting in a lower thermal conductivity when the field is perpendicular to
the heat current. This anisotropy is naturally expected to vanish at H$_{c2}$ when the vortices
overlap~\cite{maki:pr-67}. The same calculation predicts that the anisotropy of $\kappa$ will
change sign at the $T<<T_c$ limit, where the excitation of the quasiparticles perpendicular to the
vortices (Volovik effect) becomes the dominant effect and enhances the heat current in the
direction perpendicular to the magnetic field. Our data show that the anisotropy of thermal
conductivity in \Co\ is growing to the lowest temperature measured, once again indicating the need
to go to lower temperature to test the theoretical prediction. To our knowledge there are no
similar calculations for the d-wave case. More theoretical work is needed to help us understand the
magnetothermal transport in the vortex state of \Co.

\section{The superconducting phase diagram of \Co.}

\begin{figure}
\resizebox{!}{1.0\textwidth}{
\includegraphics{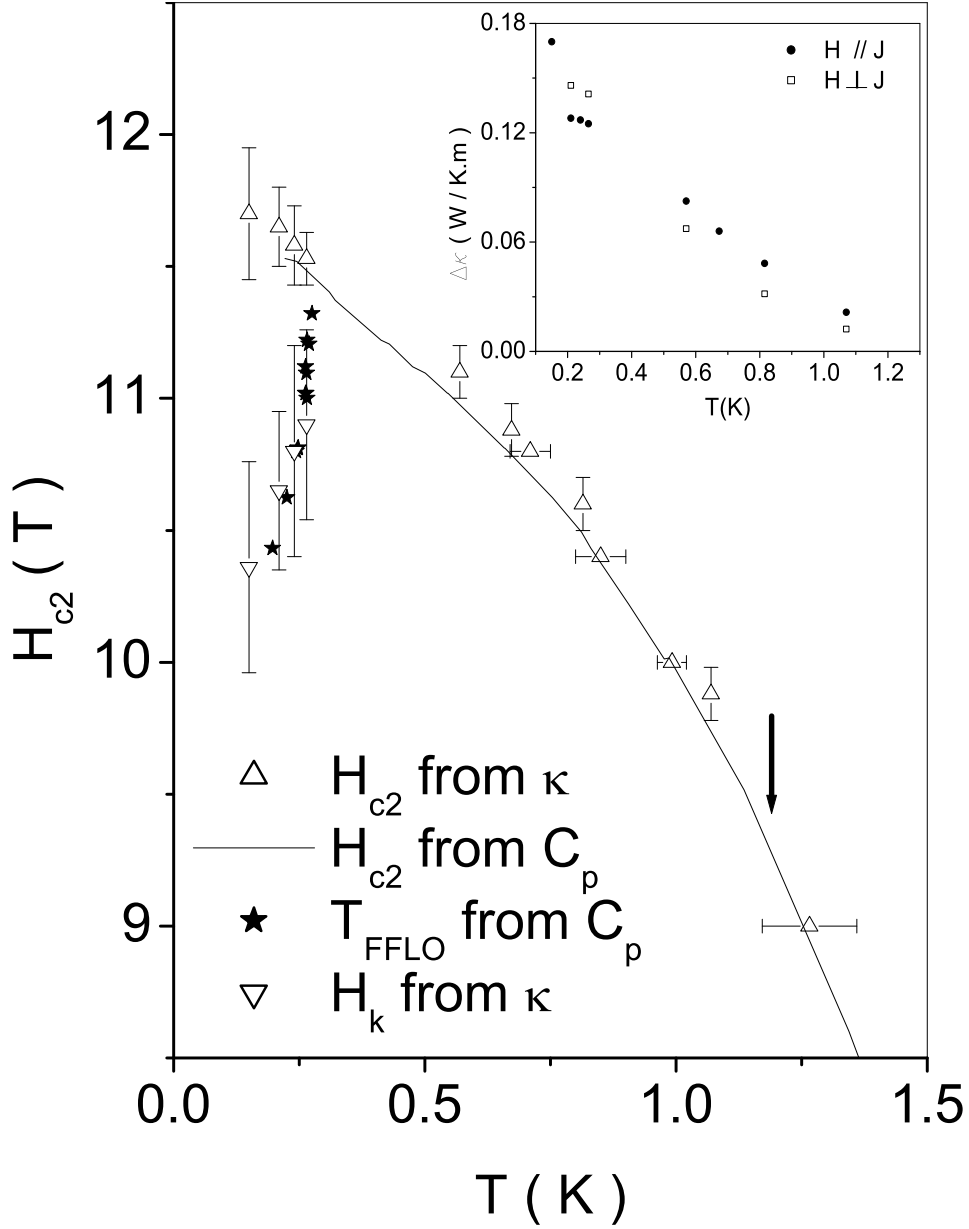}} \caption{ Magnetic field vs temperature
phase diagram of CeCoIn$_5$ deduced from thermal conductivity and specific heat measurements.
(solid line) H$_{c2}$ and ($\star$) T$_{FFLO}$ are deduced from the specific heat data of Ref [23].
 $H_{c2}$ ($\bigtriangleup$) and $H_k$ ($\bigtriangledown$) are obtained from the thermal
conductivity data (details are in the text). Arrow indicates the critical temperature where
superconducting transition changes from second to first order. Inset: the size of the jump in
$\kappa$ at the superconducting transition for the field sweeps for ($\bullet$) $H
\parallel J$ and ($\Box$) $H \perp J$.} \label{phasediag}
\end{figure}

The upper critical field H$_{c2}$, determined from both the temperature and the field sweeps, is
displayed in Fig.~\ref{phasediag} together with the critical field deduced from the specific heat.
Note that there is no difference in H$_{c2}$ for the field parallel and perpendicular to the heat
current, as expected for a tetragonal compound, since the additional in-plane anisotropy due to the
d-wave gap is zero upon the $90 \deg$ rotation. The shape of H$_{c2}$ as well as the first order
nature of the transition at low temperatures is in agreement with previous
reports~\cite{bianchi:prl-02,bianchi:prl-03a}. The strong temperature dependence of H$_{c2}$ at low
temperatures, as opposed to the saturated behavior expected in the BCS theory, is an important
observation supporting the existence of the FFLO state in CeCoIn$_5$~\cite{won:cond-mat-03}. In
fact, this is a common feature for a number of other superconductors suggested to be in the
Pauli-paramagnetic limit, such as UBe$_{13}$ \cite{radovan:condmat-03} and
$\kappa-(BEDT-TTF)_2Cu(CSN)_2$~\cite{zuo:prb-00}.

The amplitude of the jump is comparable for both field orientations and decreases almost linearly
with increasing temperature, as shown in the inset of fig.~\ref{phasediag}. This is in contrast to
the sharp decrease in the temperature step, associated with the first order transition, observed at
the critical point in magnetocaloric measurements for the field along the
c-axis~\cite{bianchi:prl-02}. It would be difficult to locate precisely the critical point where
the order of the superconducting phase transition changes from second to first, for $H
\parallel J$,because the transition occurs gradually. Nevertheless, the critical point is consistent with
that determined from specific heat measurements~\cite{bianchi:prl-03a}.

\section{Fulde-Ferrell-Larkin-Ovchinnikov transition.}

An additional feature in the $\kappa$ vs. $H$ curves displayed in Fig.~\ref{knorm-vs-H} is resolved
at the lowest temperatures. A kink appears in the data at a field $H_k$ for $H \parallel J$, and
the thermal conductivity is nearly constant between $H_k$ and $H_{c2}$ for temperature below 0.27
K. Given the sharpness of the jump at H$_{c2}$ and the first order nature of the transition, this
feature is clearly distinct from the H$_{c2}$ anomaly. This anomaly is not present in the data at
0.57 K and above, where the thermal conductivity continues to decrease up to H$_{c2}$. We therefore
interpret the region of the flat $\kappa(H)$ preceding the sharp jump at H$_{c2}$ as an enhancement
of thermal conductivity. Further, $H_k$ decreases as the temperature is reduced.
Fig.~\ref{phasediag} displays $H_k$ in the $H-T$ plane, together with the phase diagram deduced
from the specific heat measurements~\cite{bianchi:prl-03a}. The H$_k$ points coincide well with the
second low temperature phase transition line found in the specific heat below H$_{c2}$. The absence
of such a flat portion at high temperatures, where the thermal conductivity is monotonously
decreasing up to H$_{c2}$, and the good agreement of H$_k$ with the specific heat anomaly lead us
to speculate that the enhancement of thermal conductivity is due to the formation of the second low
temperature superconducting state in \Co, identified as a potential FFLO
state~\cite{bianchi:prl-03a,radovan:nature-03}. Features in the thermal conductivity  of an organic
superconductor near the superconducting critical field were also interpreted as possible signatures
of the FFLO state~\cite{tanatar:prb-02}.

The enhancement of thermal conductivity between $H_k$ and $H_{c2}$ manifests itself clearly in the
$H \parallel J$ data. It is more difficult to make a definitive statement for the data in the $H
\perp J$ geometry since data below $H_{c2}$ is rounded and a gradual rise is present in the data
for higher temperature, outside of the FFLO phase.

The enhancement of thermal conductivity for the $H \parallel J$ orientation is contrary to our
original simple-minded expectation, since in this geometry the nodal planes in the LO phase are
perpendicular to the direction of the heat current, and therefore would not be expected to enhance
thermal conductivity. There are several possible explanations of our results. Recent theoretical
work~\cite{combescot:condmat-03} suggests that the lowest energy state is not a pure LO state with
a single modulation wavevector $\vec Q$, but a modified LO state with a combination of three
modulation wavevectors. If so, one would not expect an additional anisotropy with respect to the
direction of the magnetic field due to FFLO nodal planes. The contribution to thermal conductivity
from  the nodal planes of the LO state is not a priori dominant over that from the vortices, and
must be investigated theoretically~\cite{vekhter:pc-03}. Another
scenario~\cite{boulaevskii:private-comm} suggests that the bottle neck for the heat transport along
the field direction is the vortex cores. One would expect the structure of the vortices to be
modulated by the nodal planes. The vortex cores' size might increase at the nodal planes, reducing
the bottle neck and leading to the enhancement of thermal conductivity. The interplay between the
vortex and FFLO state was theoretically studied for 2D superconductors~\cite{klein:jltp-00}. The
resulting spatial structures can be alternating nodal planes and lines of vortices, or more
intricate structures, depending on the Landau level quantization number of the order parameter.
More theoretical work on the vortex structure within the FFLO state for 3D superconductors, and its
effect on the thermal transport in particular, is called for and should clarify whether such new
structures can account for the observed enhancement of the thermal conductivity in the low
temperature superconducting state.

\section{Conclusion}
Thermal transport is a powerful probe of a superconducting state. We performed thermal conductivity
measurements to investigate the properties of both the vortex state and the second low temperature
phase of \Co, with the magnetic field applied within the $a-b$ plane of this tetragonal compound.
Our data demonstrate that the superconducting phase transition becomes first order between roughly
10 T and the superconducting critical field, in accord with previous specific heat measurements,
indicating the importance of the Pauli limiting effect in \Co. In addition, we observed a kink in
thermal conductivity for the field parallel to the direction of the heat current, coincident with
the phase transition in the second low temperature state of \Co, suggested previously to be an FFLO
state~\cite{bianchi:prl-03a,radovan:nature-03}. Thermal transport within the FFLO state at present
remains unexplored theoretically, and the observed enhancement of thermal conductivity within the
FFLO state of \Co\ is puzzling. Our experimental results present a new challenge for the
understanding of inhomogeneous superconductivity.

\section{Acknowledgements}
The authors are grateful to I.Vekhter, L.Bulaevskii, P.Fulde, F.Ronning, M.Tanatar for fruitful
discussions.



\end{document}